\documentclass{iau}
\usepackage{graphicx}

\title[IAU Symp. 289~~Advancing the physics of cosmic
distances]
{The Planetary Nebulae Luminosity Function \\ and distances to Virgo,
  Hydra~I and Coma clusters}

\author[Magda Arnaboldi \etal\ ] 
{Magda Arnaboldi$^{1,2}$, Alessia Longobardi$^3$, Ortwin Gerhard$^3$
  \and S. Okamura$^4$}

\affiliation{
  $^1$ESO, K. Schwarzschild Str. 2, 85748 Garching, Germany\\ email: {\tt marnabol@eso.org} \\[\affilskip]
  $^2$INAF, Oss. Astr. di Pino Torinese, 10025 Pino Torinese, Italy \\[\affilskip]
  $^3$Max-Planck-Institut f\"ur Extraterrestrische Physik, Postsach 1312, 85741 Garching, Germany\\[\affilskip]
  $^4$ Hosei University, Faculty of Science and Engineering Department
  of Advanced Sciences, JP Tokyo 184-8584 }

\pubyear{2013}
\volume{289}  
\pagerange{1--5}
\jname{Advancing the physics of cosmic distances}
\editors{Richard de Grijs and Giuseppe Bono, eds.}
\begin{document}

\maketitle

\begin{abstract}
  The luminosity function of planetary nebulae populations in galaxies
  within $10-15$ Mpc distance has a cut-off at bright magnitudes and a
  functional form that is observed to be invariant in different galaxy
  morphological types. Thus it is used as a secondary distance
  indicator in both early and late-type galaxies.  Recent deep surveys
  of planetary nebulae populations in brightest cluster galaxies
  (BCGs) seem to indicate that their luminosity functions deviate from
  those observed in the nearby galaxies. We discuss the evidence for
  such deviations in Virgo, and indicate which physical mechanisms may
  alter the evolution of a planetary nebula envelope and its central
  star in the halo of BCGs. We then discuss preliminary results for
  distances for the Virgo, Hydra~I and Coma clusters based on the
  observed planetary nebulae luminosity functions.  
\end{abstract}

\firstsection 

\section{Introduction}
Most of the stars in the mass range 1 to 8 $M_\odot$ go through the
planetary nebulae (PN) phase during the final stages of their lives,
before they become white dwarfs. During this phase, the nebular shell
of a PN is able to convert the UV ionizing photons into various line
emissions, from the UV to the optical and down to the NIR. Up to $15\%$
of the UV energy emitted by the central star is re-emitted in a
single line, the [OIII] $\lambda$ 5007 \AA\ line, which is the
brightest optical emission of a PN (\cite[Dopita \etal\ 1992]{Dopita+92}).

When observed in our own Milky Way galaxy, the planetary nebula's
shells and [OIII] emission are spatially resolved.  In M31 and beyond,
PNs are spatially unresolved sources of green light, thus the whole
[OIII] flux $F_{5007}$ emitted from a PN shell can be integrated and a
$m_{5007}$ magnitude is computed as (\cite[Jacoby (1989)]{Jac89}):
\begin{equation}
m_{5007} = -2.5 \log F_{5007} - 13.74
\end{equation}
For a PN population observed in external galaxies, we can then derive
the PN luminosity function (PNLF).

The PNLF was measured for PN populations in early and late-type
galaxies within $10-15$ Mpc distance (see \cite[Ciardullo \etal\
2002]{Ciardullo+2002} for a review) with the following properties:
\begin{itemize} 
\item The PNLF shows a cut-off at the bright end, whose absolute magnitude 
is $M^*$=-4.51.  
\item The shape of PNLF \underline{and} the cut-off at bright magnitudes
  are observed to be invariant in galaxies of different morphological
  types, either star-forming or quiescent, within $10-15$ Mpc distance.
\end{itemize} 
Thus the PNLF is used efficiently as a secondary distance indicator,
with some important role in the investigation of systematic biases,
because it represents one of the few methods that can be applied to
both early and late-type galaxies.

The PNLF is empirically shown to follow the analytical formula
\begin{equation}
N(m_{5007}) = C\times e^{0.307m_{5007}} \times [ 1- e^{3(m^*-m_{5007})}]
\label{PNLF}
\end{equation}
(\cite[Ciardullo \etal\ 1998]{Ciardullo+98}) where $m^*$ is the
apparent magnitude of the bright cut-off. This analytic formula
combines the observed behavior at the bright end, which is believed to
originate from the most massive, $M_{core} \simeq 0.7 M_\odot$,
surviving stellar cores (\cite[Ciardullo \etal\ 1989]{Ciardullo+89},
\cite[Marigo \etal\ 2004]{Marigo+2004}), and the slow PN fading rate
caused by the envelope expansion, at the faint end (\cite[Heinze \&
Westerlund 1963]{HenWest1963}).

It is an open question whether more physics is required to
describe PN populations in massive galaxies than what is captured by
the analytical formula $N(m_{5007})$ in eq.~\ref{PNLF}.

\section{Physics of the PNLF}\label{PhPNLF}

The theoretical basis for the $N(m_{5007})$ analytical formula is a
population of uniformly expanding, homogeneous spherical PNs ionized
by non-evolving central stars. The observed invariance of $M^*$ also
seems to indicate that the most massive surviving stellar cores all have
the same mass, regardless of the age and metallicity of the parent
stellar population. Each of these hypotheses may turn out to be
violated in different environments, and we discuss each possibility in
turn.

\underline{Constancy of $M^*$} -- A PN’s peak flux is proportional to
its core mass (\cite[Vassiliadis \& Wood 1994]{vaswood94}). The core
mass of a PN is proportional to its turnoff mass (\cite[Kalirai \etal\
2008]{kal+08}). The turnoff mass of a stellar population decreases
with age (\cite[Marigo \etal\ 2004]{Marigo+2004}). Simple stellar
population theory would then predict that the absolute magnitude of
the PNLF at the bright cut-off should become fainter already in a 1
Gyr old population, and up to four magnitude fainter in $10$ Gyrs old
stellar population.

\underline{Uniformly expanding, homogeneous spherical shells} -- The
nebular shell may not be spherical, as observed for many MW PNs, not
expanding uniformly, and not optically thick.  The presence of a hot
ISM in evolved stellar populations may also effect the mass loss
during the AGB evolution and the properties of the ionized PN shell
(\cite[Dopita \etal\ 2000]{dopmar00}, \cite[Villaver \& Stanghellini
2005]{Vilsta05}).  The interaction between nebular shell and hot gas
may also decrease the visibility lifetime of a PN, $\tau_{PN}$, which
in turns decreases the total number of PNs associated with a given
bolometric luminosity emitted the parent stellar population
(\cite[Buzzoni \etal\ 2006]{BAC2006}).

\underline{Non evolving central star} -- Simple stellar population
theory predicts low-mass cores $M_{core} \leq 0.55 M_\odot$
(\cite[Buzzoni \etal\ 2006]{BAC2006}) in an old stellar population, as the
one detected in the M87 halo (\cite[Williams \etal\
2007]{wills+07}). For such low mass cores, $\tau_{PN}$ may be shorter
than $3\times 10^4$ yrs (\cite[Buzzoni \etal\ 2006]{BAC2006}) because
the time required for the excitation of the nebular envelope
increases, and by the time it happens, the density in the nebular
shell may be too low for any significant [OIII] emission. Furthermore,
part of the stellar population with $M_{core}\sim 0.52 M_\odot$ may
omit the PN phase entirely (\cite[Bl\"ocker 1995]{bloc95}).  These
evolved stars may provide an enhanced contribution to the hotter
horizontal branch (HB) and post - HB evolution (\cite[Greggio \&
Renzini 1990]{Laura90}), as directly observed in M~32 and in the bulge
of M~31 (\cite[Brown \etal\ 1998, 2000]{Brown98}).

Physical conditions which violate the hypotheses embedded in the
analytical formula $N(m_{5007})$ may particularly occur in the halos
of BCGs galaxies at the center of massive clusters because the stellar
populations are old and immersed in a high density ICM.  We then
expect to find deviations of the observed PNLF for these systems, the
closest one being the BCG galaxy in the Virgo cluster, M87. Similar
physical conditions are expected for PN in environments like NGC~3311
in the Hydra I cluster, and the Coma cluster BCGs.

\section{The PNLF in the M87 outer halo}
The nearby clusters have a very important role in the cosmological
distance ladder. The measurement of the distance to the Virgo cluster
by the Key project was crucial for the determination of the Hubble
constant. There were several projects aimed at the measurement of the
PNLF distance to the Virgo elliptical galaxies (see \cite[Ciardullo
\etal\ 2002]{ciard+2002} for a review). The most extended PN survey in the 
M87 halo was that carried out by \cite[Ciardullo \etal\
(1998)]{Ciardullo+98} covering a $16' \times 16'$ field, centred on
M87.

The empirical PNLF measured in M87 halo from a sample of 338 PN
deviates significantly from the analytical formula $N(m_{5007})$ for a
distance modulus of $(m-M) = 30.79$ (\cite[Ciardullo \etal\
(1998)]{Ciardullo+98}). The deviations are such that {\it 1)} the PNLF
shape is different for the inner $<4'$ and the outer regions $>4'$,
with the PN candidates in the halo having brighter $m_{5007}$ than the
cut-off expected for $(m-M) = 30.79$, and {\it 2)} the PNLF for the
outer sample is not drawn from a population following eq.~\ref{PNLF},
at the 99\% level according to KS tests. At that time, the PNLF
deviations were explained in terms of a uniform Intracluster PN
(IPN) population, which extended 2~Mpc in front of the M87 halo PN
population, so that the overluminous PNs could be explained as IPNs.

\begin{figure}[b]
\begin{center}
\includegraphics[width=8cm]{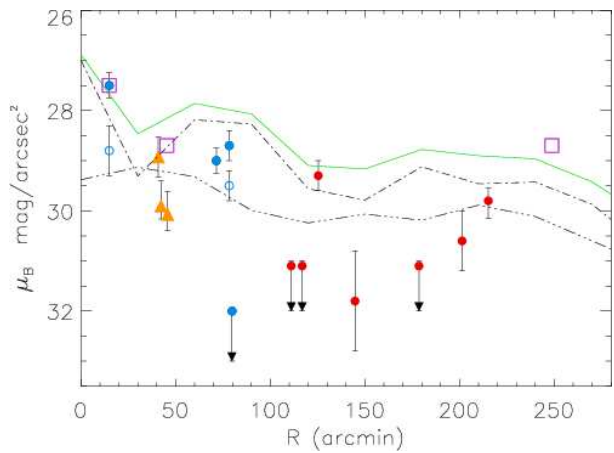} 
\caption{Surface brightness measurement of diffuse light in the Virgo
  fields (points) compared with the surface brightness profile of the
  Virgo galaxies averaged in annuli (lines); radial distances are
  computed with respect to M87. The continuous line represents the
  radial surface brightness profile from light in Virgo galaxies from
  \cite[Binggeli \etal\ 1987]{bing87}.  The dotted-dashed and double
  dotted-dashed lines correspond to the surface brightness profile
  associated with giants and dwarf galaxies, respectively. The blue
  (darker) full dots show the surface brightness measurements in the
  Virgo core.  The open circles at $10'$ and $80'$ distances indicate
  the ICL surface brightness computed from the IPNs not bound to
  galaxy halos.  The triangles represent the surface brightness of the
  ICL based on IC RGB star counts (\cite[Williams \etal\
  2007]{wills+07} and reference therein).  The red full dots at distances
  larger than $80'$ show several surface brightness measurements and
  arrows indicate their upper limits. The magenta open squares indicate the
  surface brightness average values $\mu _{\rm B}$ at 15, 50 and 240
  arcmin computed from the measurements of \cite[Feldmeier \etal\
  (2004)]{Feld04}; the measurements at $240'$ are close to M49.  The
  diffuse light is concentrated around M87 and M49, with a sharp
  decrease at a distance of $80' = 0.4$ Mpc. From
  \cite[Castro-Rodriguez \etal\ (2009)]{castro+09}.}
   \label{fig1}
\end{center}
\end{figure}

There is some tension in this proposed scenario. The spectroscopic
follow-up of PNs bound to the M87 halo showed that these PNs have a
brighter cut-off, $m^*=26.2$, than PNs in the inner regions, for which
$m^*=26.35$ (\cite[Arnaboldi \etal\ 2008]{arna+08}).  Furthermore the
extended survey carried out by \cite[Castro-Rodriguez \etal\
(2009)]{castro+09} showed that the IPN population in the Virgo cluster
is associated with its densest regions, and IPNs extend to at most 0.4
Mpc in front of M87. In Fig.\,\ref{fig1} we show the surface
brightness profile for the ICL in the Virgo cluster; it illustrates
that the ICL and IPN population are concentrated around M87 and in the
densest regions of the Virgo cluster.

In 2010 we started a new project to survey PN in M87, covering the
whole halo out to 150 kpc.  It entails an imaging survey with
SuprimeCAM@Subaru, with the aim at covering 0.5 deg$^2$ in the M87
outer halo, and a spectroscopic follow-up with FLAMES@VLT of the
selected candidates. PN candidates are identified using deep [OIII]
and off-band V images: they are selected as point like [OIII] sources
with a color excess $[OIII] - V = -1$ and no continuum
(\cite[Arnaboldi \etal\ 2002]{arna+02}).

The SuprimeCAM observations for the PN survey in M87 were taken with a
total exposure time in the narrow band [OIII] filter of about 4 hrs
and a total exposure time in the off band (V band) of about 1 hr, for
each pointing. Analysis of the data is on-going; the properties of the
PNLF from this extended PN sample, both in area and depth, are
presented in Longobardi \etal\ (2012), in prep.

\section{Detecting PN beyond Virgo}

The brightest PNs in the Hydra I cluster at 50 Mpc distance have
fluxes of $7.8\times10^{-18}$ erg~s$^{-1}$cm$^{-2}$ ($\sim 7$
photons/min on 8m tel.). In the Coma cluster, fluxes are four times
fainter. These very faint fluxes cannot be detected using a narrow
band filter because the sky noise in a $30-40$ \AA\ centred on the
redshifted [OIII] PN emission is of the same order of the signal we
want to detect. A step forward for the detection of PN in elliptical
galaxies in clusters at distances larger than 15 Mpc is the Multi Slit
Imaging Spectroscopy Technique (MSIS).

MSIS is a blind search technique, that combines a mask of parallel
multiple slits with a narrow band filter, centred on the redshifted
[OIII] 5007 \AA\ line at the Hydra I/Coma mean systemic velocity, to
obtain spectra of all PNs that lie behind the slits (\cite[Gerhard
\etal\ 2005]{ger+05}).  The sky noise at the PN emission line now
comes from a spectral range of only a few \AA, depending on slit width
and spectral resolution (\cite[Arnaboldi \etal\
2007]{arna+07}). Several tens of PNs were detected using the MSIS
observations in the Hydra~I (\cite[Ventimiglia \etal\ 2011]{ven+11})
and Coma (\cite[Gerhard \etal\ 2007]{ger+07}) clusters.

We can then use the PNLF from the MSIS data-sets to determine the
distances of Coma and Hydra~I relative to the Virgo cluster, which may
reduce systematics errors in the distance measurements to these
clusters.  This is important because several methods for the
measurements of cosmological distances (i.e. fundamental plane, high-z
SN) rely on the distance to the Coma cluster for their zero-point.

Because PNs can be observed mostly for BCGs in these clusters, we need
more advanced models for the PNLF like those computed by
\cite[M\'endez \etal\ (2008)]{M2008}, see also discussion in
Sec.~\ref{PhPNLF}. Then we need to account for the through
slit-convolution and convolution with photometric errors, plus
completeness correction; see \cite[Ventimiglia \etal\ (2011)]{ven+11}
for an overview of the procedure. Then the cumulative PNLF is a direct
distance indicator that can be used to derive the distances to
clusters beyond Virgo.

\section{Preliminary results}
Preliminary distance moduli and distances based on PNLF and MSIS samples are
\begin{itemize}
\item Virgo $(m-M)_0 \simeq 30.8$; D$_{Virgo} \simeq 15 $ Mpc
\item Hydra $(m-M)_0 \simeq 33.5$; D$_{Hydra~I} \simeq 3.43 \times
  D_{Virgo} = 51.5 $ Mpc
\item Coma $(m-M)_0 \simeq 34.9$; D$_{Coma} \simeq 6.51 \times
  D_{Virgo} = 97.7 $ Mpc
\end{itemize}

The next steps of this project include:
\begin{itemize}
\item the use of the observed PNLF in the M87 halo (Longobardi \etal\
  2012, in prep) to test models for PNLF in old stellar populations,
  including effects on AGB evolution and ionization of PN caused by
  presence of hot ISM.
\item Enlarge the MSIS sample of PN in Coma by including the PN
  samples from two additional two fields (Arnaboldi \etal\ 2013, in
  prep).
\item Carry out extensive simulation for MSIS (through
  slit-convolution, convolution with photometric errors; completeness
  correction) and error estimates on distances (Gerhard \etal\ 2013, in
  prep).  
\end{itemize}
The project of determining distances to clusters out to Coma using the PNLF is
entering into a new, exciting, phase!

\section{Acknowledgments}
MAR wishes to thank the organizers of the IAU Symposium 289 for the
opportunity to give this talk.

\end{document}